\documentstyle[aps]{revtex}
\begin{document}
\draft
\title{Measures of gravitational entropy I.\\
Self-similar spacetimes}         
\author{Nicos Pelavas and Kayll Lake\thanks{lake@astro.queensu.ca}}
\address{Department of Physics, Queen's University, Kingston, Ontario, Canada, K7L 3N6 }       
\date{\today}          
\maketitle
\begin{abstract}
We examine the possibility that the gravitational contribution to the entropy of a system can be identified with some measure of the Weyl curvature. In this paper we consider homothetically self-similar spacetimes. These are believed to play an important role in describing the asymptotic properties of more general models. By exploiting their symmetry properties we are able to impose significant restrictions on measures of the Weyl curvature which could reflect the gravitational entropy of a system. In particular, we are able to show, by way of a more general relation, that the most widely used ``dimensionless" scalar is \textit{not} a candidate for this measure along homothetic trajectories.
\end{abstract}
\section{Introduction}
Twenty years ago Penrose\cite{penrose} proposed the \textit{Weyl curvature hypothesis} (also known as the Weyl tensor hypothesis) which requires that the Weyl tensor is zero at the big bang\cite{Rothman}. It was expected that this condition would imply a subsequent evolution close to a Friedman-Robertson-Walker (FRW) model. The hypothesis is motivated by the need for a low entropy constraint on the initial state of the universe when the matter content was in thermal equilibrium. Penrose has argued that the low entropy constraint follows from the existence of the second law of thermodynamics, and that the low entropy in the gravitational field is tied to constraints on the Weyl curvature. Presumably, this constraint is a consequence of quantum gravity.

There is evidence that the requirement that the Weyl tensor is \textit{zero} at the big bang is too strong. For example, in polytropic perfect fluid spacetimes Anguige and Tod \cite{Anguige} have recently provided uniqueness results that show that if the Weyl tensor is zero at the big bang, the spacetime geometry must be \textit{exactly} FRW in a neighbourhood of the big-bang in accord with the FRW conjecture\cite{Scott}. Nonetheless, the motivation for some form of the Weyl curvature hypothesis remains. From a thermodynamic point of view, whereas the big bang would normally be considered a state of maximum entropy (assuming approximate thermodynamic equilibrium), there is every indication that the entropy of the universe is
\textit{increasing}. It is only natural to expect that this apparent paradox may reflect the omission of gravitational degrees of freedom when in fact the gravitational entropy of the big bang was, in some sense, low. Increasing gravitational entropy would naturally be reflected by increasing local anisotropy, and the Weyl tensor reflects this. As a consequence, the identification of a measure of the Weyl curvature which reflects the gravitational contribution to the entropy may be considered a fundamental step towards a deeper understanding of general relativity even if the Weyl tensor wasn't exactly zero initially. To put it differently, whereas spacetime is usually defined as a connected time-oriented four-dimensional Lorentz manifold, in practice time-oriented is weakened to time-orientable \cite{Oneill}. We do not have a fundamental construction which determines the future direction in spacetime. A purely gravitational entropy would provide such a construction. Whereas one might argue that gravitational entropy is restricted to some average in cosmological spacetimes, neither the averaging procedure nor the concept of ``cosmological" is clear.

In an attempt to be somewhat more precise, to begin consider a scalar field $ \textit{P}\; (x^{a}) $, constructed from the Riemann tensor and its covariant derivatives (equivalently the metric, its inverse and partial derivatives to arbitrary order), and a (smooth) timelike trajectory $x^a(\tau)$ with (unit) tangent $u^a$. We call $\textit{P}\; (x^{a})$ a measure of ``gravitational entropy" ( alternatively the ``gravitational epoch") if it is monotone along timelike trajectories while providing some measure of local anisotropy. The ``future" is determined by the choice
\begin{equation}
\frac{\rm d \textit{P}}{\rm d \tau} > 0. \label{epoch1}
\end{equation}
We take the view that there could be several distinct fields $ \textit{P}\; (x^{a}) $ with a claim to be the ``gravitational entropy". Then again, it may be that no suitable $ \textit{P}\; (x^{a}) $ exists.
The search for a gravitational arrow of time, within this formulation, reduces to a search for the scalar $ \textit{P}\; (x^{a}) $. This search can be simplified in a number of ways, including restrictions on: \textit{(i)} spacetimes for which (\ref{epoch1}) is expected to hold, \textit{(ii)} the trajectories $x^a(\tau)$ along which the evolution of $ \textit{P}\; (x^{a}) $ is allowed to be measured, and \textit{(iii)} the construction of $\textit{P}\; (x^{a})$. Here we start with the view that a purely gravitational entropy should exist in all non-conformally flat spacetimes. This is the only restriction 
we place on \textit{(i)}.  

It is clear that some restriction of type \textit{(ii)} is necessary since, for example, no development of a purely gravitational entropy can be expected along the Killing trajectories of a stationary spacetime. It is the purpose of this paper to explore how this restriction generalizes to homothetic trajectories in self-similar spacetimes.  As a special 
case of a more general result, we will show how the evolution of a scalar field $\textit{P}\; (x^{a})$ behaves along homothetic trajectories in a self-similar spacetime.  In particular we show that any ``dimensionless" scalar (as defined below) has no development.

As regards restriction \textit{(iii)}, a first step is to construct $\textit{P}$ directly in terms of elementary scalars derivable from the Riemann tensor without differentiation. The relation between gravitational entropy and anisotropy could be reflected by (\ref{epoch1}) if the Weyl tensor, in some sense, dominated $ \textit{P}$. Indeed, the simplest choice is \cite{fn1}
\begin{equation}
\textit{P} = \textit{C}_{\alpha \beta}^{\; \; \; \; \gamma \delta} \textit{C}_{\gamma \delta}^{ \; \; \; \; \alpha \beta} \label{weyl1}
\end{equation} 
where $\textit{C}_{\alpha \beta \gamma \delta}$ is the Weyl tensor. 
 It is known that this choice for $ \textit{P}$ fails at \textit{isotropic singularities} \cite{Wainwright} and fails to handle both decaying and growing perturbation modes \cite{Rothman1}. The ``dimensionless" scalar
\begin{equation}
\textit{P} = \frac{\rm \textit{C}_{\alpha \beta}^{\; \; \; \; \gamma \delta} \textit{C}_{\gamma \delta}^{ \; \; \; \; \alpha \beta}}{\rm \textit{R}_{\alpha }^{ \beta}\textit{R}_{\beta}^{ \alpha}}, \label{weyl3}
\end{equation}
where $\textit{R}_{\alpha \beta}$ is the Ricci tensor, satisfies (\ref{epoch1}) near an isotropic singularity \cite{Wainwright} and produces the correct dependence for both growing and decaying modes \cite{Rothman1}. However, Bonnor \cite{Bonnor} has also argued that (\ref{weyl3}) fails to give the correct sense of time for a radiating source. 
 
In this paper we show, for example,  that (\ref{weyl3}) is not an acceptable candidate for gravitational entropy along the homothetic trajectories of any self-similar spacetime. Nor indeed is any ``dimensionless" scalar. Our arguments exploit the symmetry properties of homothetically self-similar spacetimes. These calculations are of physical interest since self-similar spacetimes are very widely studied \cite{Carrandcoley} and are, for example, believed to play an important role in describing the asymptotic properties of more general models. The paper is organized as follows: After a review of background material, we present a general result from which our investigation of invariants follows. This is followed by a discussion 
regarding restrictions and alternative choices for the construction of a gravitational entropy. 

\section{Background}
A four dimensional manifold \textit{M}, with a metric \textit{g} admits a homothetic motion if there
exists a global vector field $\xi$, such that 
\begin{equation}
\L_{\xi}g_{\alpha\beta}=2{\phi}g_{\alpha\beta}, \label{homothetic}
\end{equation}
where $\L$ is the Lie derivative and $\phi=\nabla_{\alpha}\xi^{\alpha}$ is some nonzero constant \cite{Yano} \cite{fn2}. Here we call a spacetime \textit{self-similar} if the vector field $\xi$ is timelike. We are interested in invariants ($I$) constructed from the Riemann tensor and its covariant derivatives (equivalently the metric, its inverse and partial derivatives to arbitrary order). It is convenient to build these out of the trace-free Ricci tensor ($S^\alpha_\beta \equiv R^\alpha_\beta - \frac{\rm 1}{\rm 4}\delta^\alpha_\beta R$), Weyl tensor ($C_{\alpha\beta}\,^{\gamma\delta}$), and the dual Weyl ($C^{\ast}_{\alpha\beta}\,^{\gamma\delta}$) tensor \cite{CR}. We introduce two definitions: The \emph{degree} of an invariant $I$ is the number of multiplications that are involved in its construction, and, the \emph{order} of an invariant $I$ is the largest number of derivatives contained in one of its factors.

Since $\L$ and the covariant derivative $\nabla_{\alpha}$ commute $\it{iff}$ $\L_{\xi}{\Gamma}_{\alpha\beta}^{\gamma}=0$ \cite{Yano}, and since homothetic motions are a subset of affine collineations, 
we can interchange the Lie derivative and the covariant derivative freely.  
Conversely $\L$ and $\nabla^{\alpha}$ do \emph{not} commute, an extra term arises. Therefore the Lie derivative of a differential invariant involving $\nabla^{\alpha}$ along a homothetic vector field will have extra terms with respect to 
its non-differential counterpart (see below).
From (\ref{homothetic}) it follows that
\begin{equation}
\L_{\xi}S_{\alpha}^{\beta}=-2{\phi}S_{\alpha}^{\beta}, \label{S}
\end{equation}
\begin{equation}
\L_{\xi}C_{\alpha\beta}\,^{\gamma\delta}=-2{\phi}C_{\alpha\beta}\,^{\gamma\delta}, \label{C}
\end{equation}
and
\begin{equation}
\L_{\xi}C^{\ast}_{\alpha\beta}\,^{\gamma\delta}=-2{\phi}C^{\ast}_{\alpha\beta}\,^{\gamma\delta}. \label{Cstar}  
\end{equation}

If there exists a homothetic vector field $\xi$  then it follows that any tensor $T^{a_{1} \cdots a_{i}}\,_{a_{i+1} \cdots a_{j}}$ constructed from the metric tensor via covariant/contravariant derivatives, partial derivatives, and/or contractions will necessarily have the form
\begin{equation}
\L_{\xi}(T^{a_{1} \cdots a_{i}}\,_{a_{i+1} \cdots a_{j}})=\kappa{\phi}T^{a_{1} \cdots a_{i}}\,_{a_{i+1} \cdots a_{j}}, \label{arbtensor}
\end{equation}
where $\kappa$ is an even integer (see below).
As a consequence, every time an index is lowered
\begin{equation}                      
 \kappa \mapsto{\kappa+2}, \label{lowering}
\end{equation}  
and every time an index is raised               
\begin{equation}
\kappa \mapsto{\kappa-2}. \label{raising}
\end{equation}
Since invariants $I$  are characterized by having a conservation of covariant and contravariant indices, whenever an index is lowered on an invariant one must necessarily be raised.  Therefore the effect of (\ref{lowering}) and (\ref{raising}) is nullified, and so $\kappa$ is well defined for $I$. Moreover, this observation implies that when dealing with the Lie derivative of invariants along homothetic vector fields, without any loss of generality one can choose to forget about the indices. For example, using (\ref{lowering}) and (\ref{raising}) one can immediately write from (\ref{C}):
\begin{equation}
\L_{\xi}C_{\alpha\beta\gamma}\,^{\delta}=0,\ \L_{\xi}C_{\alpha\beta\gamma\delta}=2{\phi}C_{\alpha\beta\gamma\delta},\ 
\L_{\xi}C^{\alpha}\,_{\beta\gamma\delta}=0, \ldots  
\end{equation}
The Lie derivative of a tensor with $i$ contravariant indices and $j$ covariant indices 
will have the same coefficient $\kappa$, irrespective of the labels of the indices, and the position of the indices.

Consider a tensor $T=T^{a_{1} \cdots a_{u_{i}}}\,_{a_{u_{i}+1} \cdots a_{n}}$ with $u_{i}$ 
contravariant indices and $d_{i}=n-u_{i}$ covariant indices, 
such that $\L_{\xi}(T)=\kappa\phi T$. Then, different index configurations will map
\begin{equation}
\kappa \mapsto \kappa\prime=\kappa+2(u_{i}-u_{i}\prime) \label{kappaprime} 
\end{equation}
where the number of contravariant indices maps from $u_{i} \mapsto u_{i}\prime$.
Using (\ref{kappaprime}) one can simply write the possible range of $\kappa\prime$ as
\begin{equation}
\kappa - 2d_{i}\leq \kappa\prime \leq \kappa+2u_{i}. \label{range}
\end{equation}
The non commutativity of $\L$ and $\nabla^{\alpha}$ 
is illustrated if we consider some tensor $T$ of arbitrary rank. Then
\begin{equation}
\L_{\xi}(\nabla^{\alpha}T)=\nabla^{\alpha}\L_{\xi}(T)-2{\phi}\nabla^{\alpha}T. \label{noncommutativity} 
\end{equation}
Generalizing to order $m$, with $j$ contravariant derivatives and $m-j$ covariant derivatives we have,
\begin{equation}
\L_{\xi}(\nabla^{a_{1} \cdots a_{j}}\,_{a_{j+1} \cdots a_{m}}T)=
\nabla^{a_{1} \cdots a_{j}}\,_{a_{j+1} \cdots a_{m}}\L_{\xi}(T)
-2j\phi\nabla^{a_{1} \cdots a_{j}}\,_{a_{j+1} \cdots a_{m}}T.  \label{deriv}
\end{equation}
Using (\ref{deriv}), we have the following result:
\begin{equation}
\L_{\xi}(T)=2\kappa\phi T, 
\end{equation}
for some natural number $\kappa$, if and only if 
\begin{equation}
\L_{\xi}(\nabla^{a_{1} \cdots a_{\kappa}}\,_{a_{\kappa+1} \cdots a_{m}}T)=0. \label{lcovariant}
\end{equation}
Note that the number of covariant derivatives in (\ref{lcovariant}) is irrelevant.
Again, since homothetic motions are a subset of affine collineations, we can write a corresponding relationship
to (\ref{deriv}) for partial derivatives, by simply replacing the covariant/contravariant derivatives with covariant/contravariant 
partial derivatives. This gives:
\begin{equation}
\L_{\xi}(\partial^{a_{1} \cdots a_{j}}\,_{a_{j+1}\cdots a_{m}}T)=
\partial^{a_{1} \cdots a_{j}}\,_{a_{j+1} \cdots a_{m}}\L_{\xi}(T)
-2j\phi\partial^{a_{1} \cdots a_{j}}\,_{a_{j+1} \cdots a_{m}}T. \label{partial}
\end{equation}
If $T$ is the metric tensor then 
(\ref{deriv}) and (\ref{partial}) reduce to 
\begin{equation}
\L_{\xi}(\nabla^{a_{1} \cdots a_{j}}\,_{a_{j+1} \cdots a_{m}}g_{\alpha\beta})=
2(1-j)\phi \nabla^{a_{1} \cdots a_{j}}\,_{a_{j+1} \cdots a_{m}}g_{\alpha\beta}, \label{metricderiv}
\end{equation}
and
\begin{equation}
\L_{\xi}(\partial^{a_{1} \cdots a_{j}}\,_{a_{j+1}\cdots a_{m}}g_{\alpha\beta})=
2(1-j)\phi \partial^{a_{1} \cdots a_{j}}\,_{a_{j+1} \cdots a_{m}}g_{\alpha\beta}. \label{metricpartial} 
\end{equation}
Using (\ref{metricderiv}) and (\ref{metricpartial}), we have the further result that $\kappa$ in (\ref{arbtensor}) is a an even integer. This raises the obvious question: Does there exist an index configuration such that $\kappa\prime=0$?  Using (\ref{kappaprime}), we set $\kappa\prime=0$, and quickly arrive at the conclusion: If the number of contravariant indicies is mapped to $u_j^{\prime} = \kappa/2+u_{j}$, and $0 \leq u_{j}^{\prime} \leq n$ then $\kappa^{\prime}=0$. Therefore with this particular index configuration, $\L_{\xi}(T)=0$.
\section{General Case}
Consider a tensor $T=\nabla^{m_{1}}T_{1}^{p_{1}} 
\cdots \nabla^{m_{n}}T_{n}^{p_{n}}$ of arbitrary rank, and define a factor of $T$ to be 
$\nabla^{m_{i}}T_{i}^{p_{i}}$, hence $T$ has $n$ factors and the index $i$ characterizes the 
$i^{th}$ factor of $T$. Let $\nabla^{m_{i}}$ denote $m_{i}$ derivatives, with $\kappa_{i}$ contravariant derivatives, and $m_{i}-\kappa_{i}$ covariant derivatives \textit{i.e.} $\nabla^{m_{i}}\equiv \nabla^{a_{1_{i}} \cdots a_{\kappa_{i}}}\,_{a_{(\kappa+1)_{i}} \cdots a_{m_{i}}}$. Define $T_{i}^{p_{i}}=T_{1_{i}} \cdots T_{q_{i}} \cdots T_{p_{i}}$, so that $T_{i}^{p_{i}}$ denotes  $p_{i}$ tensors multiplied together.  A factor of $T_{i}^{p_{i}}$ is some tensor $T_{q_{i}}$.  Associated to every $T_{q_{i}}$ is an integer $n_{q_{i}}$, denoting the coefficient in the Lie
derivative of the tensor $T_{q_{i}}$ \textit{i.e.} $L_{\xi}(T_{q_{i}})=-2n_{q_{i}}{\phi}T_{q_{i}}$. We obtain the following result: \\

Given a homothetic vector field $\xi$, and a tensor $T=\nabla^{m_{1}}T_{1}^{p_{1}} 
\cdots \nabla^{m_{n}}T_{n}^{p_{n}}$ of arbitrary rank, then
\begin{equation}
L_{\xi}(T)=-2\left[\sum_{i=1}^{n}\left(\kappa_{i}+\scriptstyle{\sum}_{q_{i}=1_{i}}^{p_{i}}\textstyle 
n_{q_{i}}\right)\right]{\phi}T. \label{central} 
\end{equation}\\

This can be shown as follows:
Since $T_{i}^{p_{i}}$ represents $p_{i}$ tensors multiplied together (possibly all different), we can write
$$T_{i}^{p_{i}}=T_{1_{i}} \cdots T_{p_{i}}$$ \mbox{so that $L_{\xi}(T_{i}^{p_{i}})=L_{\xi}(T_{1_{i}} \cdots T_{p_{i}})$.}
Now if for some $q_{i}$, \mbox{$1_{i} \leq q_{i} \leq p_{i}$} we have $$L_{\xi}(T_{q_{i}})=-2n_{q_{i}}\phi T_{q_{i}}$$ and so
$$L_{\xi}(T_{i}^{p_{i}})=-2\left(\sum_{q_{i}=1_{i}}^{p_{i}}n_{q_{i}}\right)\phi T_{i}^{p_{i}}.$$
For the tensor $T=\nabla^{m_{1}}T_{1}^{p_{1}} \cdots \nabla^{m_{n}}T_{n}^{p_{n}}$ we have,
\begin{eqnarray}
      L_{\xi}(T)=L_{\xi}(\nabla^{m_{1}}T_{1}^{p_{1}})\nabla^{m_{2}}T_{2}^{p_{2}} \cdots
 \nabla^{m_{n}}T_{n}^{p_{n}}+ \cdots \nonumber \\
\mbox{} \cdots + \nabla^{m_{1}}T_{1}^{p_{1}} \cdots L_{\xi}(\nabla^{m_{n}}T_{n}^{p_{n}}) 
\end{eqnarray}
\begin{eqnarray}
        =(\nabla^{m_{1}}L_{\xi}(T_{1}^{p_{1}})-2\kappa_{1}\phi\nabla^{m_{1}}T_{1}^{p_{1}})\nabla^{m_{2}}T_{2}^{p_{2}} \cdots 
\nabla^{m_{n}}T_{n}^{p_{n}} + \cdots \nonumber \\
\mbox{} \cdots + \nabla^{m_{1}}T_{1}^{p_{1}} \cdots (\nabla^{m_{n}}L_{\xi}(T_{n}^{p_{n}})-2\kappa_{n}\phi\nabla^{m_{n}}T_{n}^{p_{n}})    
\end{eqnarray}
\begin{eqnarray}
=\nabla^{m_{1}}L_{\xi}(T_{1}^{p_{1}})\nabla^{m_{2}}T_{2}^{p_{2}} \cdots \nabla^{m_{n}}T_{n}^{p_{n}}+ \cdots     
+ \nabla^{m_{1}}T_{1}^{p_{1}} 
\cdots \nabla^{m_{n}}L_{\xi}(T_{n}^{p_{n}})  \nonumber \\
\mbox{} - 2\left(\sum_{i=1}^{n} \kappa_{i}\right)\phi\nabla^{m_{1}}T_{1}^{p_{1}}
\cdots \nabla^{m_{n}}T_{n}^{p_{n}} 
\end{eqnarray}
\begin{equation}
      =-2\left[\sum_{i=1}^{n}\sum_{q_{i}=1_{i}}^{p_{i}}n_{q_{i}}\right]\phi\nabla^{m_{1}}T_{1}^{p_{1}} \cdots \nabla^{m_{n}}T_{n}^{p_{n}}-
2\left[\sum_{i=1}^{n}\kappa_{i}\right]\phi\nabla^{m_{1}}T_{1}^{p_{1}} \cdots \nabla^{m_{n}}T_{n}^{p_{n}} 
\end{equation}
\begin{equation}
      =-2\left[\sum_{i=1}^{n}\left[\kappa_{i}+\scriptstyle{\sum}_{q_{i}=1_{i}}^{p_{i}}\textstyle{n_{q_{i}}}\right]\right]\phi\nabla^{m_{1}}T_{1}^{p_{1}} 
\cdots \nabla^{m_{n}}T_{n}^{p_{n}}, 
\end{equation}
from which we obtain (\ref{central}).

\section{Invariants}
Consider an invariant $I$ that is built out of the Weyl tensor and/or dual Weyl tensor, such that $\L_{\xi}(I)=\kappa\phi I$. 
Then $I$ provides a measure of local anisotropy and is monotone along timelike homothetic trajectories $\xi$.
It therefore is a candidate for providing a measure of gravitational entropy in self-similar spacetimes.

By taking the $trace(T)$ in (\ref{central}), (\ref{central}) specializes to an invariant built out of an arbitrary number of covariant and/or contravariant derivatives, with tensors of arbitrary rank. Given an invariant $I$, (\ref{central}) reduces to:
\begin{equation}
L_{\xi}(I)=-2\left[\sum_{i=1}^{n}\left(\kappa_{i}+\scriptstyle{\sum}_{q_{i}=1_{i}}^{p_{i}}\textstyle n_{q_{i}}\right)\right]{\phi}I. \label{invariant}
\end{equation} 
Now suppose that $T_{i}^{p_{i}}$ has $u_{i}$ contravariant indices, and $d_{i}$ covariant indices.  With the same notation as above, it follows from a simple counting argument that 
\begin{equation}
\sum_{i=1}^{n}\left[2\kappa_{i}+u_{i}-d_{i}-m_{i}\right]=0 \label{invconstraint}
\end{equation}
for $I$ where $\kappa_{i}\leq m_{i}$. Relation (\ref{invconstraint}) provides a constraint on the possible combinations of order (i.e. derivatives) and degree, that are allowable in order to construct an invariant.  Suppose that $\forall$ $i$ $u_{i}+d_{i}=S_{i}$ is even\cite{fn3}, and $\forall$ $i$ 
the number of derivatives, $m_{i}$, is odd\cite{fn4}. Then a consequence of (\ref{invconstraint}) is that there does not exist an invariant $I$, with both odd order and odd degree.

As a special case, suppose that there exists an index configuration in which $n_{q_{i}}=1\ \forall q_{i}$. This is the case when $T_{q_{i}}\epsilon\{S_{\alpha}\,^{\beta}, C_{\alpha\beta}\,^{\gamma\delta}, C^{\ast}_{\alpha\beta}\,^{\gamma\delta}\}$.
It follows that
\begin{equation}
L_{\xi}(I)=-2\left[\sum_{i=1}^{n}\left(\kappa_{i}+p_{i}\right)\right]{\phi}I. \label{kp}
\end{equation}
Even though (\ref{kp}) was obtained by setting $n_{q_{i}}=1$, (\ref{kp}) will also hold if the weaker condition
$\sum_{q_{i}=1}^{p_{i}}\textstyle n_{q_{i}}=p_{i}$ is satisfied.  This is analogous to raising an index on $T_{q_{j}}$, then lowering one on $T_{q_{k}}$, so that $T_{q_{i}}$ does not necessarily have to be sampled from 
$\{S_{\alpha}\,^{\beta}, C_{\alpha\beta}\,^{\gamma\delta}, C^{\ast}_{\alpha\beta}\,^{\gamma\delta}\}$ 
with this particular index configuration. Nevertheless, specializing further by setting $\sum_{i=1}^{n}\textstyle u_{i}-d_{i}=0$, which happens to be the case for 
$\{S_{\alpha}\,^{\beta}, C_{\alpha\beta}\,^{\gamma\delta}, C^{\ast}_{\alpha\beta}\,^{\gamma\delta}\}$\cite{fn5}
 since $u_{i}=d_{i}\ \forall i$, if $m_{i}=m\ \forall\ i$ 
i.e. all factors have the same number of derivatives, we have $\sum_{i=1}^{n}\kappa_{i}=mn/2$ from (\ref{invconstraint}) so that
\begin{equation}
L_{\xi}(I)=-\left[mn+2\sum_{i=1}^{n}p_{i}\right]{\phi}I. \label{mp}
\end{equation}
 If we also set $p_{i}=1\ \forall\ i$, i.e. there exist only one tensor in each factor of the invariant, then
\begin{equation}
L_{\xi}(I)=-np{\phi}I,\ \ p=m+2. \label{mn}
\end{equation}
Equation (\ref{mn}) applies to invariants of degree $n$ and order $p$. For example, an invariant of degree 2 and order 3 is 
$\nabla^{\alpha}S_{\beta\gamma}\nabla^{\beta}S_{\alpha}^{\gamma}$.
We have $p=m+2$, where $m$ is the number of derivatives, both contravariant and covariant and $2$ comes from the fact that 
$\kappa$ in (\ref{arbtensor}) is even and also that $n_i=p_i=1$. For invariants built out of $\{S_{\alpha}\,^{\beta}, C_{\alpha\beta}\,^{\gamma\delta}, C^{\ast}_{\alpha\beta}\,^{\gamma\delta}\}$, the 2 coincidentally matches the number of derivatives of the metric required to build these tensors. 
In index-free notation,  write $I=\nabla^{m}T_{1} \cdots \nabla^{m}T_{n}$, where $\nabla^{m}$ denotes $\kappa_{i}$ contravariant derivatives and 
$m-\kappa_{i}$ covariant derivatives, for the $i^{th}$ factor, and
$T_{i}$ is some arbitrary rank tensor with $L_{\xi}(T_{i})=-2n_{i}{\phi}T_{i}$. If we let  
$N=\sum_{i=1}^{n} n_{i}$ ($N$ is an invariant under the operations of raising or lowering indices as in (\ref{raising}) and (\ref{lowering})),
then an invariant of degree $n$ and order $p$ satisfies
\begin{equation}
L_{\xi}(I)=-\left[2N+mn\right]\phi I. \label{Nm}
\end{equation}
Observe that if in (\ref{Nm}) we take $T_{i}{\epsilon}\{S, C, C^{\ast}\}$ $\forall$ $i$ where this notation is meant to 
indicate that the indices on trace-free Ricci, Weyl and dual Weyl tensors are irrelevant, then $N=\sum_{i=1}^{n} n_{i}=n$, and so  (\ref{Nm}) reduces to (\ref{mn}).
Equation (\ref{mn}) holds for all differential and non-differential invariants built from the tensors $\{S, C, C^{\ast}\}$.

Specializing further, consider non-differential invariants, that is $p=2$. Then equation (\ref{mn}) simplifies to
\begin{equation}
L_{\xi}(I)=-2n{\phi}I.   \label{inv}
\end{equation}
Equation (\ref{inv}) will hold for all non-differential invariants built from $\{S, C, C^{\ast}\}$. Define the \textit{dimension }$d=np$. We have the useful result:

Given invariants $I_{1}$ and $I_{2}$ built from $\{S, C, C^{\ast}\}$ and of orders $p_{1}$ and $p_{2}$ and degrees ${\mathit{n_{1}}}$ and $\mathbf{\mathit{n_{2}}}$ respectively,
 \begin{equation}
L_{\xi}\left(\frac{I_{1}}{I_{2}}\right)=-\left(d_{1}-d_{2}\right)\phi\frac{I_{1}}{I_{2}}. \label{corollary}
\end{equation}

Relation(\ref{corollary}) motivates an equivalence relation among invariants: $I_{1} \sim I_{2}$ if 
$d_{1}=d_{2}$, so all invariants of the same dimension are in the same equivalence class. Two invariants from the same equivalence class satisfy
\begin{equation}
L_{\xi}\left(\frac{I_{1}}{I_{2}}\right)=0,\label{equivalence}
\end{equation}
so that $I_{1}$ and $I_{2}$ are simply proportional along the homotheticities. For example, if $I_{1}=\nabla^{2}T$ and $I_{2}=T^{2}$ then $\nabla^{2}T=cT^{2}$, $c=constant$, along the integral curves defined by $\xi$. The notation used to represent the invariant $I_{1}$ is meant to indicate that it is constructed from two contravariant/covariant derivatives of some tensor T. Similarly, the invariant $I_{2}$ is constructed from the full contraction of two tensors which we denote by $T^2 = T_{1}T_{2}$. The indicies are omitted here, since they are irrelevant.

A table illustrating the equivalence classes for $L_{\xi}(I)=-d\phi I=-np\phi I$ is given below:
\begin{center}
\begin{tabular}{||c|c|c|c||} \hline\hline
$d \epsilon {N}$ & Order ${p}$ & Degree $n$ & Invariant \\ \hline
2 & 2 & 1 & $T_{1}$ \\ \hline
4 & 2 & 2 & $T_{1}T_{2}=T^{2}$ \\
4 & 4 & 1 & $\nabla_{1}\nabla_{2}T=\nabla^{2}T$ \\ \hline
6 & 2 & 3 & $T_{1}T_{2}T_{3}=T^{3}$ \\
6 & 3 & 2 & $\nabla_{1}T_{1}\nabla_{2}T_{2}$ \\
6 & 6 & 1 & $\nabla^{4}T$ \\ \hline
8 & 2 & 4 & $T^{4}$ \\
8 & 4 & 2 & $\nabla^{2}T\nabla^{2}T$ \\
8 & 8 & 1 & $\nabla^{6}T$ \\
\vdots & \vdots & \vdots & \vdots \\ \hline \hline      
\end{tabular}
\end{center}
In the table $T_{1}$ represents an invariant (i.e. the trace of a tensor) built out of second order derivatives of the metric tensor.  $T^{2}=T_{1}T_{2}$ represents an invariant built out of the contraction of two tensors $T_{1}$ and $T_{2}$ 
constructed from second order derivatives of the metric tensor. For example, $S_{\alpha}^{\beta}S_{\beta}^{\alpha}$. The notation $\nabla_{1}\nabla_{2}T$ is meant to indicate two distinct contravariant/covariant derivatives of some tensor $T$.  (The derivatives could be contracted together or with  the tensor $T$.) An example of $\nabla^{2}T=\nabla_{1}\nabla_{2}T$ would be $\nabla^{\alpha}\nabla^{\beta}S_{\alpha\beta}$. Since the dimensions of $S_{\alpha}^{\beta}S_{\beta}^{\alpha}$ and $\nabla^{\alpha}\nabla^{\beta}S_{\alpha\beta}$ are equivalent, then they are in the same equivalence class as shown.

It is clear from the table that the ratio of any invariants, from the same equivalence class , say $I_{1}/I_{2}$,
will satisfy (\ref{equivalence}). This implies that $I_{1}/I_{2}$ is constant along timelike homothetic trajectories.  
Therefore $I_{1}/I_{2}$ cannot be a candidate for a measure of gravitational entropy in self-similar spacetimes.
Conversely, if $I_{1}$ and $I_{2}$ come from different equivalence classes and $I_{1}/I_{2}$ is built out of the Weyl 
tensor and/or dual Weyl tensor, then this ratio may be a candidate for providing a measure of gravitational entropy 
in a self-similar spacetime.  

\section{Discussion}
The scalar (\ref{weyl3}) is the simplest example of what we have called a dimensionless scalar, since the invariant
in the numerator $I_{1} = \textit{C}_{\alpha \beta}^{\; \; \; \; \gamma \delta} \textit{C}_{\gamma \delta}^{ \; \; \; \; \alpha \beta}$, and the invariant in the denominator $I_{2} = \textit{R}_{\alpha}^{\beta}
\textit{R}_{\beta}^{ \alpha}$ have dimensions of $d_{1} = d_{2} = 4$. As a consequence, (\ref{weyl3}) does not provide a measure of gravitational entropy along homotheticities. The options available for the further study of gravitational entropy include:
\textit{i)} The exclusion of self-similar spacetimes, \textit{ii)} The exclusion of homothetic trajectories for the measurement of $ \textit{P}\; (x^{a}) $, \textit{iii)} The study of scalars $ \textit{P}\; (x^{a}) $ which have dimension, and \textit{iv)} The study of objects more general than scalars for the construction of $ \textit{P}\; (x^{a}) $.

We take the view that \textit{i)} is untenable (see \textit{e.g.} \cite{Carrandcoley}). As regards \textit{ii)}, and by reference to stationary spacetimes ($\phi = 0$), it may well be that (\ref{epoch1}) holds only for some timelike trajectories and in the limit along at least one. The present calculation shows how simple the evolution of $ \textit{P}\; (x^{a}) $ is along homotheticities, even if $ \textit{P}\; (x^{a}) $ has dimension. Whereas the number of possibilities for $ \textit{P}\; (x^{a}) $ at order $p=2$ is limited (there are but 4 independent Ricci and 4 independent Weyl invariants), invariants of order $p > 2$ offer many more possibilities. By \textit{iv)} we wish to point out that pure scalars (those derivable from the metric, its inverse and partial derivatives to arbitrary order) are not the only possibility since, for example, tangent fields could be used in the construction of $ \textit{P}\; (x^{a}) $. \\

This work was supported in part by a grant to K.L. from the Natural Sciences and Engineering Research Council of Canada.
Portions of this work made use of GRTensorII \cite{grt}.


\begin{references}
\bibitem{penrose} See, for example, R. Penrose 1979, in \emph{General Relativity: an Einstein
centennial volume}~ed. W. Israel and S. W. Hawking (Cambridge: CUP. An updated readable account is given in R. Penrose 1990, \emph{The Emperor's New Mind} (Oxford: OUP).
\bibitem{Rothman} The Weyl curvature hypothesis has been interpreted in a number of ways, but the form proposed by Penrose is made clear in a comment by Penrose in T. Rothman ``A Phase Space Approach to Gravitational Entropy" (gr-qc/9906002), South African Relativistic Cosmology Symposium, Feb 1999 (To appear in GRG).
\bibitem{Anguige} K. Anguige and K. P. Tod 1999 \textit{Annals Phys.} \textbf{276 } 257-293.
\bibitem{Scott}S. M. Scott and G. Ericksson 1998 \textit{Proc. Int. Sem. Math. Cosmol., Potsdam 1998} Editors: M. Rainer, H.-J. Schmidt, (WSPC: Singapore).
\bibitem{Oneill} See, for example, B. O'Neill 1983, \textit{Semi-Riemannian Geometry} (Academic Press: New York).
\bibitem{fn1} We use classical index notation throughout. This is done for readability. The calculations reported here are, for the most part, far more easily executed with spinors.
\bibitem{Wainwright} See, for example, S. W. Goode, A. A. Coley and J. Wainwright 1992 \textit{Class. Quantum Grav.} \textbf{9} 445  and references therein.
\bibitem{Rothman1} T. Rothman and P. Anninos Phys. Rev. D \textbf{55} 1948 (1997).
\bibitem{Bonnor} W. B. Bonnor Phys. Lett. \textbf{A 122} 305 (1987).
\bibitem{Carrandcoley} B. J. Carr and A. A. Coley 1999 Class. \textit{Quantum Grav.} \textbf{16} R13 and references therein.
\bibitem{Yano}K. Yano 1955 \emph{The Theory of Lie Derivatives and its Applications} (North-Holland: Amsterdam).
\bibitem{fn2} Setting $\phi=0$ in what follows shows the evolution along the Killing trajectories of a stationary spacetime.
\bibitem{CR} See, for example, R. Penrose and W. Rindler 1984 \textit{Spinors and spacetime Vol. 1} (Cambridge: CUP)
\bibitem{fn3} The total number of indices in $T_{i}^{p_{i}}$ is even, this applies to invariants built out of trace-free Ricci, Weyl and conjugate Weyl tensors.
\bibitem{fn4}This can be relaxed to just demanding that $\sum_{i=1}^{n}m_{i}$ is odd.
\bibitem{fn5}We choose $T_{q_{i}}\epsilon
\{S_{\alpha}\,^{\beta}, C_{\alpha\beta}\,^{\gamma\delta}, C^{\ast}_{\alpha\beta}\,^{\gamma\delta}\}$ because it allows a simplification
of the coefficient $\sum_{i=1}^{n}\left(\kappa_{i}+\scriptstyle{\sum}_{q_{i}=1}^{p_{i}}\textstyle n_{q_{i}}\right)$, but note that this 
coefficient will also hold for other index configurations, i.e. choosing $T_{q_{i}}\epsilon\{S, C, C^{\ast}\}$ without regard for the configuration of the indicies.
\bibitem{grt}GRTensorII is a package which runs within MapleV. It is entirely distinct from packages distributed with MapleV and must be obtained independently. The GRTensorII software and documentation is distributed freely on the
World-Wide-Web from the address {\tt http://grtensor.phy.queensu.ca}
\end{references}
\end{document}